\RequirePackage{luatex85}

\documentclass[a4paper,
        keeplastbox,  
        ]{jacow}
\usepackage{nidanfloat}
\usepackage{pdfpages,multirow,ragged2e} %
\makeatletter%
	\ifboolexpr{bool{xetex}}
	 {\renewcommand{\Gin@extensions}{.pdf,%
	          .png,.jpg,.bmp,.pict,.tif,.psd,.mac,.sga,.tga,.gif,%
	          .eps,.ps,%
	          }}{}
\makeatother

\ifboolexpr{bool{xetex} or bool{luatex}} 
 {}                   
 {\usepackage[utf8]{inputenc}}      

\usepackage[USenglish]{babel}
\usepackage[colorlinks=true,urlcolor=blue,linkcolor=blue,citecolor=blue]{hyperref}

\ifboolexpr{bool{jacowbiblatex}}%
 {%
 \addbibresource{jacow-test.bib}
 \addbibresource{biblatex-examples.bib}
 }{}
\begin{document}

\title{The feedback system for the longitudinal coupled-bunch instabilities in the J-PARC Main Ring}

\author{Yasuyuki Sugiyama\thanks{yasuyuki.sugiyama@kek.jp}, Masahito Yoshii, KEK/J-PARC, Tokai, Ibaraki, Japan. \\
		Fumihiko Tamura\textsuperscript{1}, JAEA/J-PARC, Tokai, Ibaraki, Japan.}
	
\maketitle

\begin{abstract}
The J-PARC Main Ring (MR) has achieved the delivery of the 30 GeV proton beam with the beam power of 500 kW to the neutrino experiment in May 2018.
The longitudinal coupled-bunch instabilities have been observed in the MR for the beam power beyond 470 kW.
Since more significant coupled-bunch oscillation was observed in higher beam power,
the mitigation of the coupled-bunch instabilities is necessary for the stable beam delivery with the beam power beyond 500 kW.
A new feedback system was developed to suppress the coupled-bunch instabilities.
The feedback system consists of a wall current monitor, an FPGA-based feedback processor, RF power amplifiers, and an RF cavity as a longitudinal kicker. 
The synchrotron sideband components of the beam signal picked up by the wall current monitor are detected by the feedback processor and used for the feedback control.
The single-sideband filtering technique is employed in the feedback processor to control each coupled-bunch mode separately.
To accommodate the change of the synchrotron frequency during the acceleration, a synchrotron frequency tracking CIC filter is used as a low pass filter in the single-sideband filter.
We present the preliminary beam test results to suppress the beam oscillation with the developed feedback system.
\end{abstract}

\bibliographystyle{IEEEtran}

\section{Introduction}
\subsection{J-PARC MR}

\begin{figure}[hbtp]
\centering
\includegraphics[width=.9\linewidth]{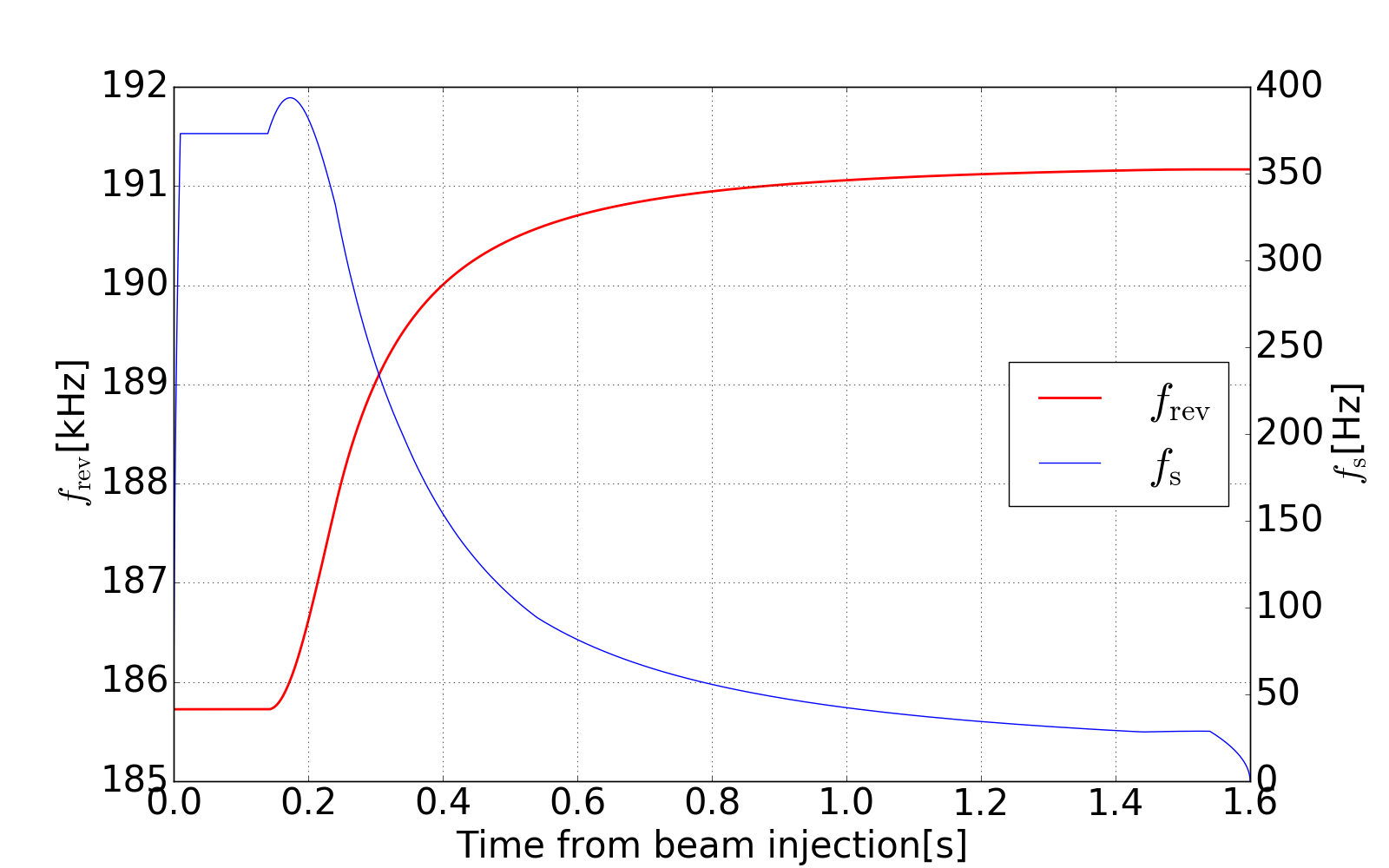}
\caption{The revolution frequency $f_\mathrm{rev}$ and the synchrotron frequency $f_\mathrm{s}$ in the J-PARC MR from the injection to the extraction.\cite{Sugiyama2018a}}
\label{fig:frev}
\end{figure}

%
The
Main Ring synchrotron (MR)\cite{Koseki2012} in the
Japan Proton Accelerator Research Complex
(J-PARC)\cite{Nagamiya2012}
is a high intensity proton synchrotron which accelerates protons from 3 GeV to 30 GeV. 
The MR delivers the proton beams to the neutrino experiment 
by the fast extraction (FX).
%
The parameters of the MR and its RF system for the FX are shown in Table \ref{table:mrparam}.
Figure \ref{fig:frev}
 shows the revolution frequency, $f_\mathrm{rev}$, and the synchrotron frequency, $f_\mathrm{s}$, in the MR. 
During the acceleration from 3 GeV to 30 GeV,
the synchrotron frequency is changing largely from 350 Hz at the injection to 30 Hz at the extraction along with the change of the revolution frequency from 185 kHz to 191 kHz.
%
%
\begin{table}[hbt]
\centering
\caption{Parameters of the J-PARC MR and its RF system for the FX.}
\begin{tabular}{ll}
\hline
parameter&value\\
\hline
circumference& 1567.5 m\\
energy& 3--30 GeV\\
beam intensity& (achieved) $2.6\times10^{14}$ ppp\\
beam power& (achieved) 500 kW\\
repetition period& 2.48 s\\
accelerating period& 1.4 s\\
accelerating frequency $f_\mathrm{RF}$& 1.67--1.72 MHz\\
revolution frequency $f_\mathrm{rev}$& 185--191 kHz\\
harmonic number $h_\mathrm{RF}$& 9\\
number of bunches $N_b$& 8\\
maximum rf voltage& 320 kV\\
No. of cavities& 7 (h=9), 2 (h=18)\\
Q-value of rf cavity& 22\\
\hline
\end{tabular}
\label{table:mrparam}
\end{table}

%

The MR delivers the 30 GeV proton beam with the beam power of 500 kW,
which corresponds to $2.6\times10^{14}$ protons per pulse in every 2.48 s,
to the neutrino experiment in May 2018.
During studies toward higher beam intensity, the longitudinal bunch oscillation is appeared to be an issue to achieve higher beam intensities than 500 kW,
and it is necessary to suppress the oscillation for stable beam acceleration at the beam power higher than 500 kW.
\begin{figure}[h]
\centering
\includegraphics[width=.9\linewidth]{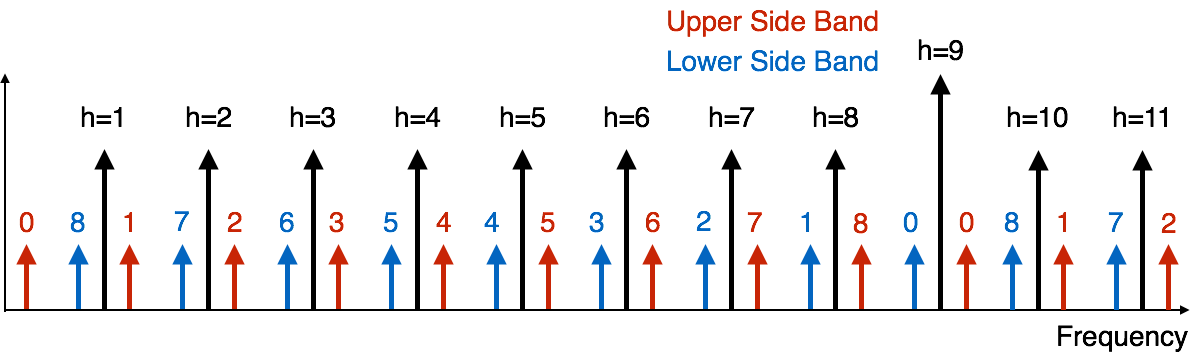}
\caption{The spectra of the synchrotron sidebands corresponding to the coupled bunch oscillation mode for the J-PARC MR\cite{Sugiyama:IBIC2017-TUPCF10}.}
\label{fig:HandModes}
\end{figure}

\subsection{Coupled Bunch Oscillation}
For M bunches, there are M modes of the CB oscillation with the mode number $n=0 ... M-1$.
The CB modes can 
be seen in the spectrum of the beam signal as synchrotron sidebands of the harmonic components\cite{Pedersen1977}.
The CB modes appear as the Upper synchrotron Side Bands (USBs) and
the Lower synchrotron Side Bands (LSBs). 
Below the accelerating frequency,
the LSB and the USB with the CB mode $n$ can be expressed as follows:
\begin{eqnarray}
f_n^{\mathrm{USB}}&=&nf_\mathrm{rev}+mf_s\\
f_n^{\mathrm{LSB}}&=&(M-n)f_\mathrm{rev}-mf_s,
\label{fn}
\end{eqnarray}
where 
$f_\mathrm{rev}$ is the revolution frequency,
$f_s$ the synchrotron frequency,
and $m$ the type of the synchrotron motion.
The case with $m=1$ corresponds to the dipole oscillation, and the case with $m=2$ corresponds to the quadruple oscillation.

There are 9 CB modes for the MR since the harmonic number of the MR is 9.
The spectra of synchrotron sidebands corresponding to the CB modes in the MR up to the harmonic $h=11$ are illustrated in Fig. \ref{fig:HandModes}.


Based on the analysis of the longitudinal CB oscillation
\cite{Sugiyama:IBIC2017-TUPCF10},
strong CB oscillation of mode $n=8$ was observed in the harmonic component of $h=8,10$.
The suppression of CB oscillation in this mode is a key to achieve the beam power higher than 500 kW.

\section{the longitudinal mode-by-mode feedback system}
\begin{figure}[htbp]
\centering
\includegraphics[width=.9\linewidth]{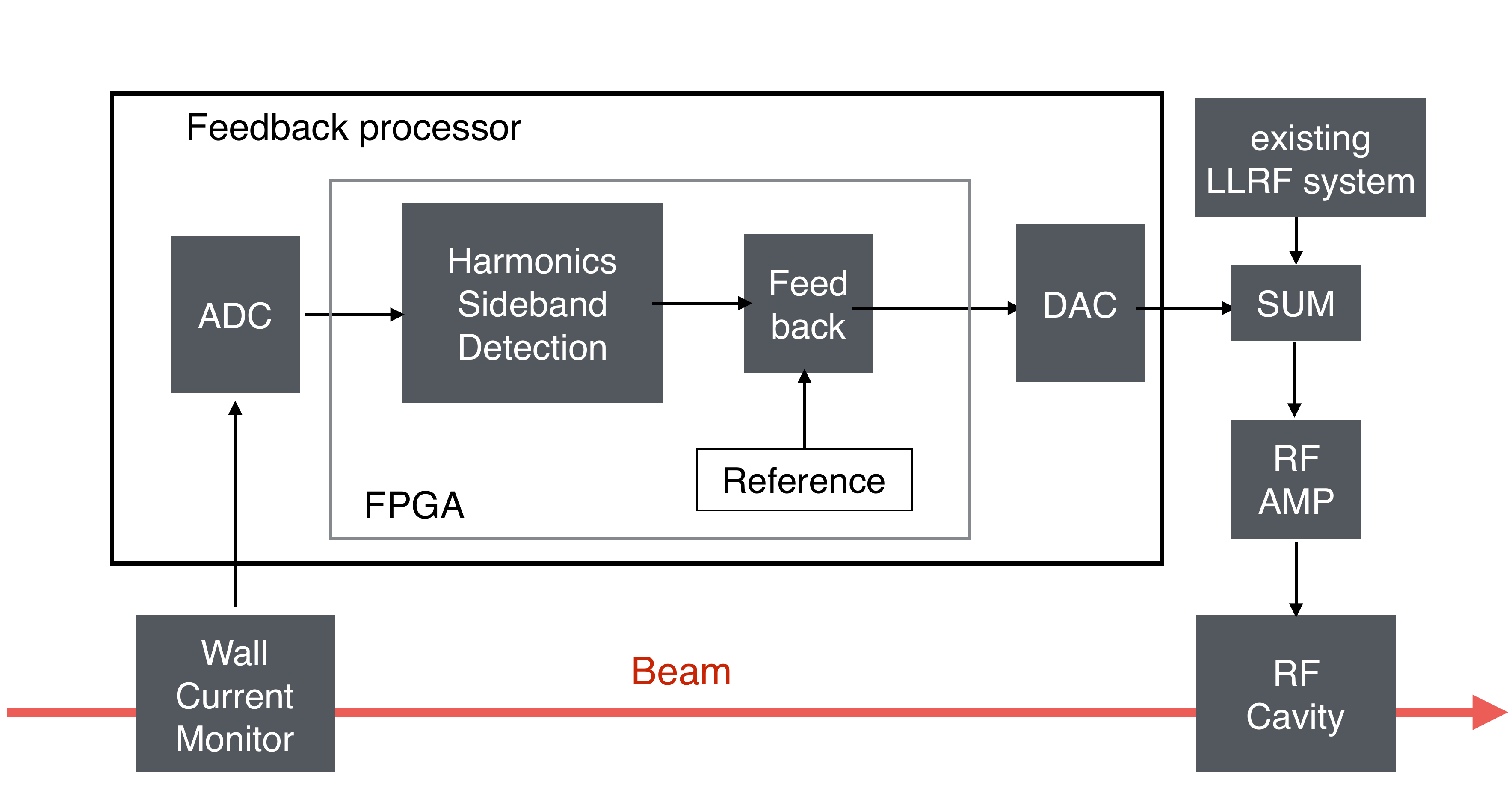}
\caption{Block diagram of the longitudinal mode-by-mode feedback system\cite{Sugiyama2018a}.}%
\label{fig:FBsystem}
\end{figure}
\begin{figure}[htbp]
\centering
\includegraphics[width=.8\linewidth]{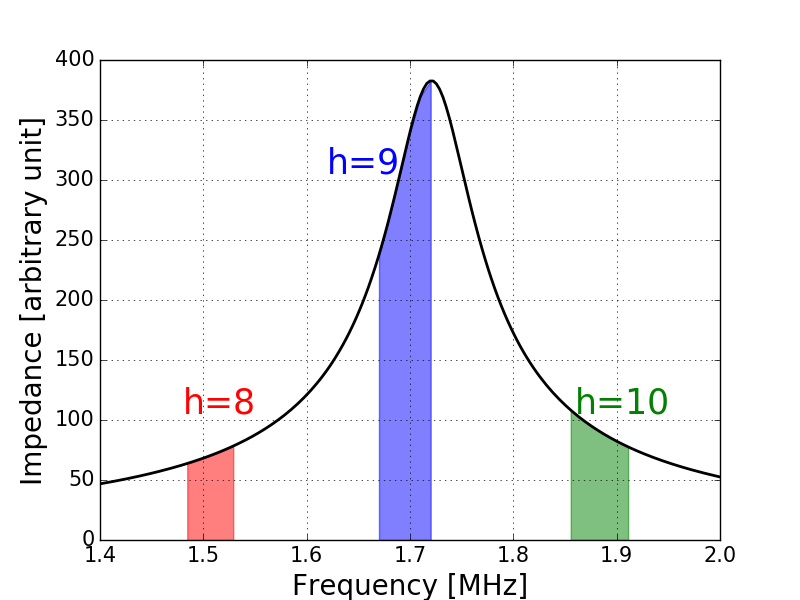}
\caption{The frequency response of the impedance of the RF cavity\cite{Sugiyama2018a}. }
\label{fig:impedance}
\end{figure}
%
We developed a longitudinal mode-by-mode feedback system to mitigate CB instabilities.

Figure~\ref{fig:FBsystem} shows the block diagram of the longitudinal mode-by-mode feedback system.
The feedback system consists of a Wall Current Monitor (WCM), an FPGA-based feedback processor, RF power amplifiers, and an RF cavity as a longitudinal kicker. 
The beam signal is detected by a WCM and fed to a feedback processor.

The CB oscillation components of the beam signal are detected by the feedback processor and used it for the feedback control.
The feedback signal from the feedback processor is led to a high-level RF (HLRF) system consisting of power amplifiers and an RF cavity.


The RF cavity used for the acceleration in the MR is used as a longitudinal kicker in the feedback system.
Figure \ref{fig:impedance} shows the frequency response of the impedance of the RF cavity used for the acceleration in the MR.
The RF cavity has impedance large enough to generate the kick voltage for the feedback in the frequency range for $h=8,10$ component.

Since the existing RF cavity is used as a longitudinal kicker in the feedback system,
the feedback system utilizes the existing HLRF system used in the MR.
The feedback signal from the feedback processor is summed with the RF signal from the current low-level RF (LLRF) system for the acceleration\cite{Tamura2013}.
The summed signal is amplified by the RF power amplifiers and fed into the 
 RF cavity. 

\section{Feedback processor}

We developed an FPGA-based longitudinal mode-by-mode feedback processor for the feedback system.
The feedback processor was manufactured by Mitsubishi Electric TOKKI Systems Corporation based on the MicroTCA.4 architecture.
\begin{figure}[h]
\centering
\includegraphics[width=.8\linewidth]{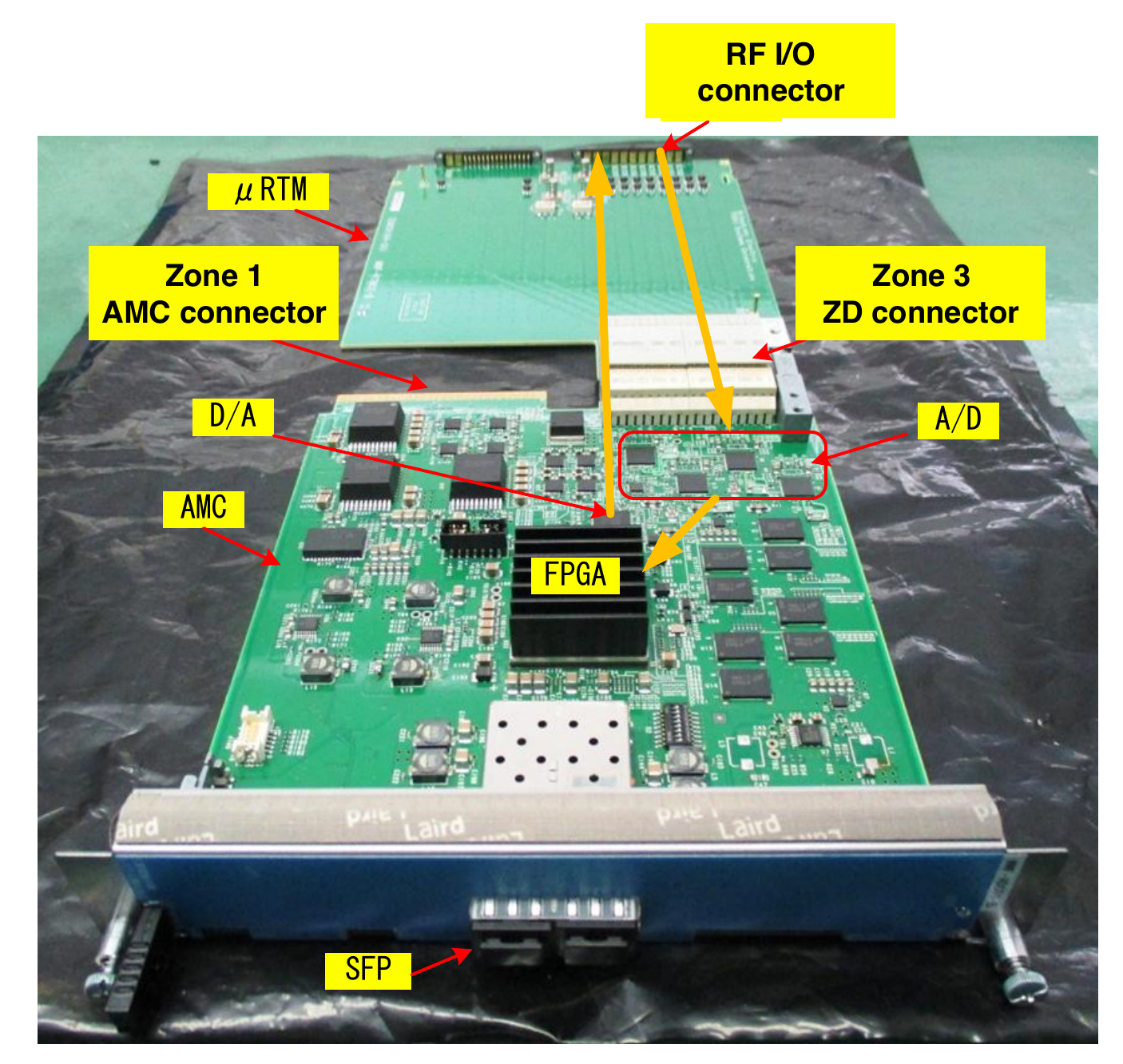}
\caption{Pictures of the longitudinal mode-by-mode feedback processor.\cite{Ryoshi2015}}
\label{fig:FBprocessor_photo}
\end{figure}

Figure \ref{fig:FBprocessor_photo} shows a picture of the developed feedback processor.
The feedback processor consists of an Advanced Mezzanine Card (AMC) and a Rear Transition Module (RTM).
%
%
The general purpose AMC module\cite{Ryoshi2015} developed by Mitsubishi Electric TOKKI Systems Corporation is used in the system.
The AMC has 8 ADC channels by 4 ADC\footnote{Texas Instruments ADC16DX370 16-bit 370-MSPS ADC} chips, 2 DAC channel with a DAC\footnote{Analog Devices AD9783 16-bit 500-MSPS DAC} chip, and an FPGA.
A Xilinx Zynq XC7Z045 SoC FPGA is used as an FPGA on the AMC module. 
In addition to signal processing,
an EPICS IOC is implemented in the embedded Linux on the Zynq FPGA and used for remote management of the module.
%
The RTM is used as the signal transition module for the AMC.

\begin{figure*}[tb]
\centering
\includegraphics[width=.8\linewidth]{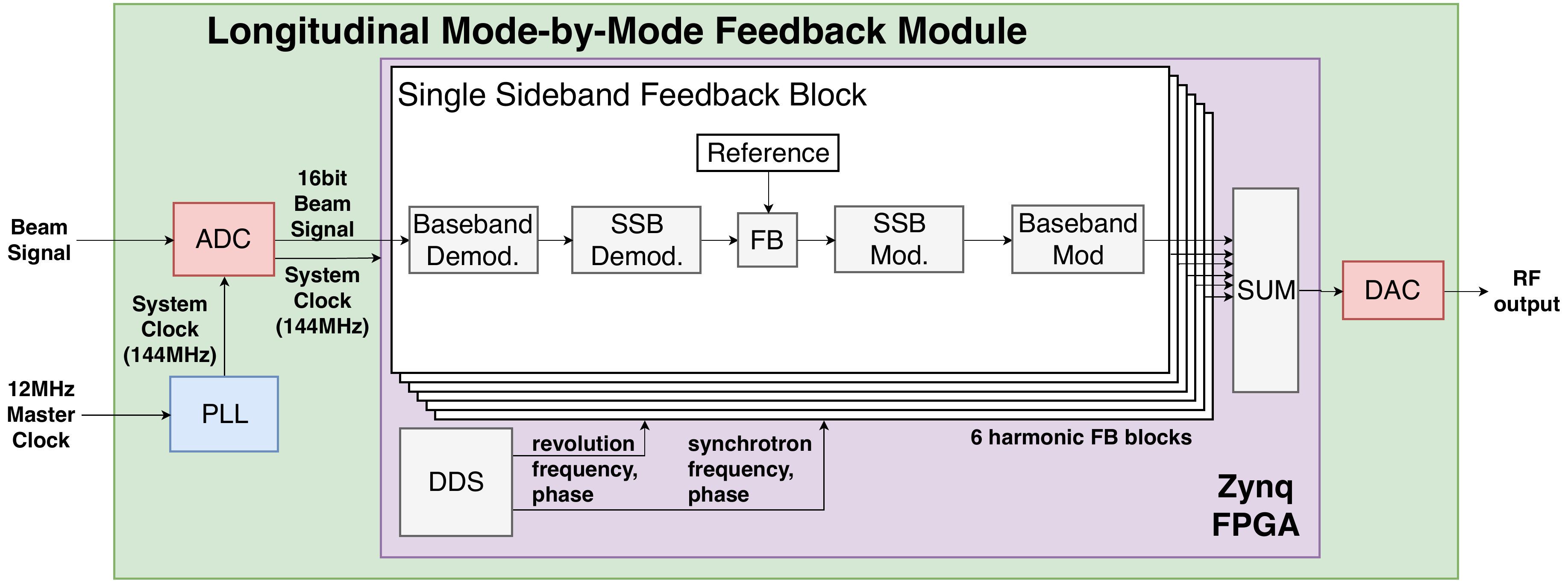}
\caption{Block diagram of the longitudinal mode-by-mode feedback processor.}
\label{fig:FBprocessor}
\end{figure*}

Figure~\ref{fig:FBprocessor}
%
%
%
 shows the block diagram of the longitudinal mode-by-mode feedback processor.
The feedback processor works at a system clock of 144 MHz from the RTM module.
The frequency and the phase signals are generated in the Direct Digital Synthesis (DDS) block.

The digitized waveform of the beam signal is converted into the baseband I/Q signal of each harmonic component by a baseband modulator.
The synchrotron sidebands for each CB mode are detected by a single sideband demodulator using the single-sideband filter (SSBF)\cite{Kriegbaum1977} technique.
The control of each CB mode is achieved by the feedback control of each synchrotron sideband component filtered by the SSBF. 
The outputs of the feedback control are converted to the RF signal by the single sideband modulator and the baseband modulator.
The RF signals are summed and converted to the analog signal by the DAC.
The feedback processor can process six different harmonic components at the same time.
\subsection{Single Sideband Feedback}
The LSBs and USBs of the synchrotron sidebands in the harmonic component are detected separately by the SSBF to control the oscillation of each CB mode.


\begin{figure*}[tb]
\centering
\includegraphics[width=.9\linewidth]{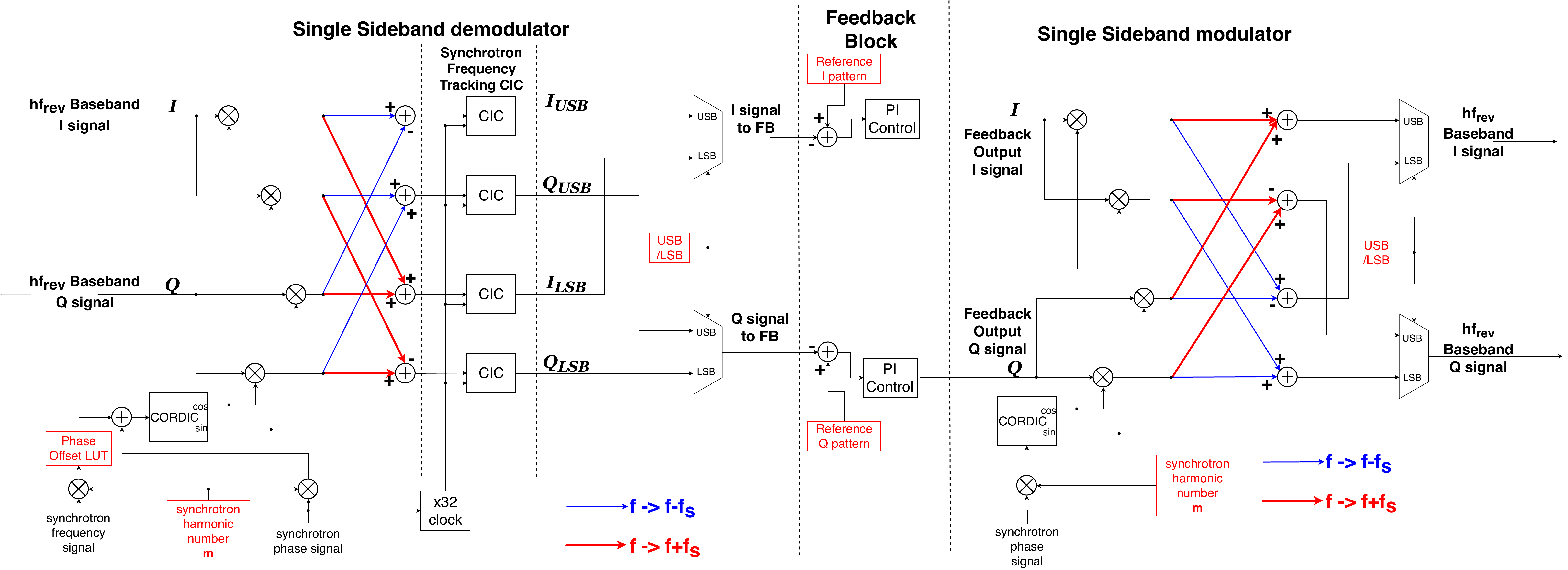}
\caption{Block diagram of the Single Sideband feedback block.}
\label{fig:SSBfilter}
\end{figure*}

Figure \ref{fig:SSBfilter} shows the block diagram of the single-sideband feedback block implemented in the feedback processor.


In the baseband I/Q signal of the harmonic component,
the synchrotron sidebands located at $f=\pm mf_s$. 
By mixing the baseband I/Q signals with the sine and cosine signal of $f=\pm mf_s$,
the spectra are shifted so that the components of $f=\pm mf_s$ locate at $f=0$ after summation and subtraction.
At the same time, $f=0,\mp mf_s$ components are shifted to the sideband with $f=\mp mf_s, 2\mp mf_s$, respectively.


The I/Q signals of the LSB and USB are detected by applying a narrow Low Pass Filter (LPF) to the mixed I/Q signals.
The suppression of the baseband signal and unwanted sidebands at the LPF is key to detecting and controlling only the selected sideband signal.
To accommodate the change in the synchrotron frequency during the cycle,
we selected a 2-stage frequency tracking CIC filter as a narrow LPF in the SSBF.
The frequency tracking CIC filter proposed by J.C.~Molendijk\cite{Molendijk2017} is a CIC filter that changes its notch position along with the frequency pattern.
By setting the synchrotron frequency pattern as the first notch frequency of the filter,
the filter can always suppress the baseband and unwanted sideband components while they change their positions during the acceleration.
The USB and LSB signals are detected separately at the SSBF, and the signal of the selected sideband is used as the input signal to the feedback block.

For the feedback control, a Proportional (P) and an Integral (I) controller is implemented in the feedback logic.

After the feedback control,
the output I/Q signals from the feedback block are modulated to the baseband I/Q signal of the harmonic component after mixing with the sine and cosine signal of $f=\pm mf_s$.

The sine and the cosine signals for mixing are generated by the CORDIC.
The phase offset to the CORDIC can be set separately for the LSB and USB to compensate the phase frequency response of the system.
The Look Up Table (LUT) for the phase offset is addressed by the harmonics of the synchrotron frequency. 

\section{Preliminary test results}
The feedback performance of the developed feedback system was tested with the beam.
In the beam test,
one of the accelerating RF cavities was used as a longitudinal kicker for the feedback system.
The cavity used for the feedback system is only used as a longitudinal kicker without the accelerating RF signal from the existing LLRF system.
The beam loading compensation by RF feedforward method\cite{Tamura2013} was disabled for the RF cavity. 

\subsection{Beam Excitation}
The beam oscillation excited by the feedback system was used as a controlled oscillation to test the response and the performance of the feedback system.

\begin{figure}[t]
\centering
\includegraphics[width=.8\linewidth]{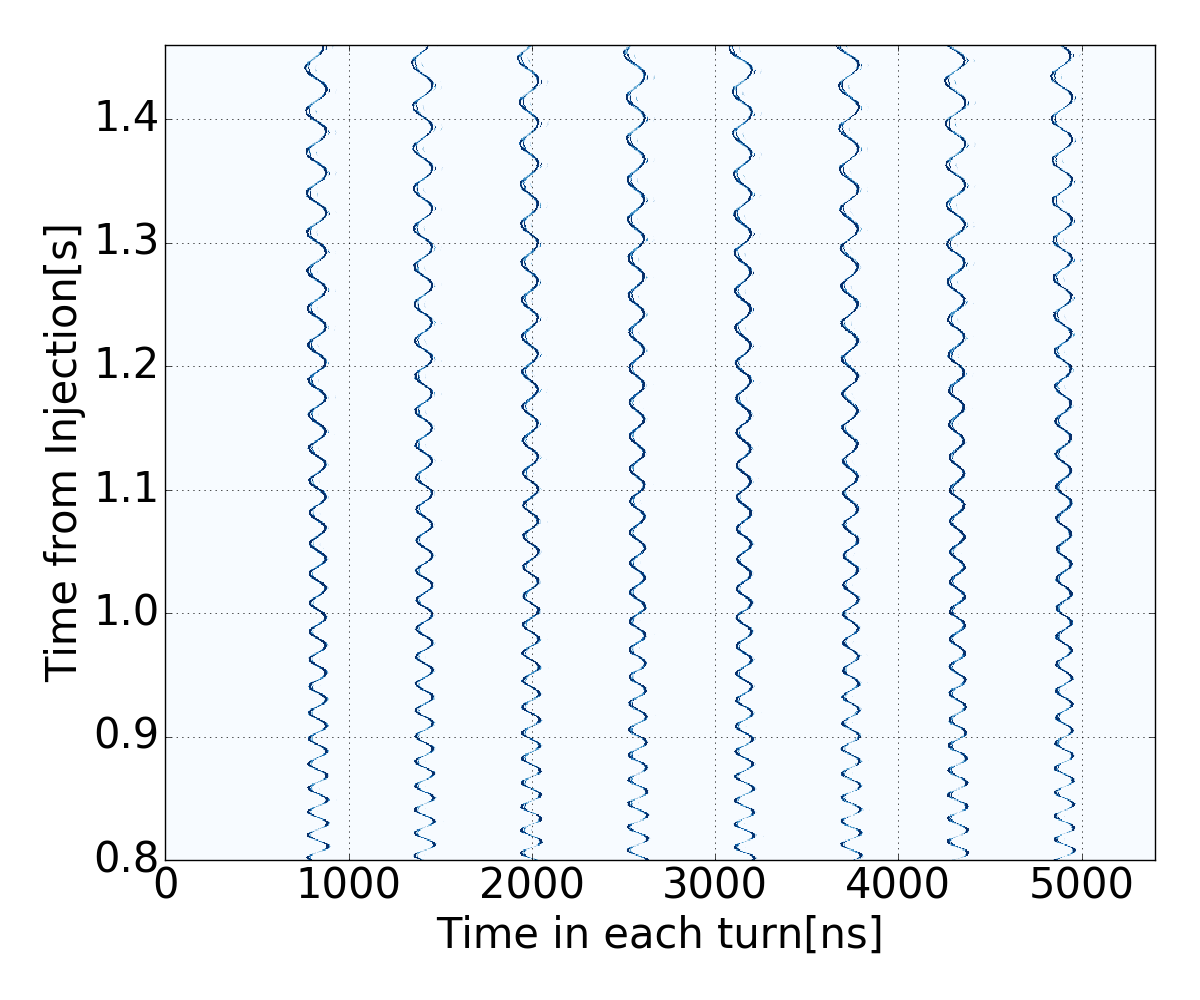}
\caption{The mountain plot for the beam excitation measurement.}
\label{fig:mountain_excite}
\end{figure}

The excitation measurement was done with the 12-kW beam low enough to get the stable beam without any longitudinal oscillation.
The beam excitation was done by setting the non-zero set point to the reference IQ pattern in the feedback module.
%
The USB of the harmonic component of $h=8$ was excited to have the CB oscillation of mode $n=8$. 
The calculated synchrotron frequency pattern was used for the sideband detection and also for the modulation in the feedback processor. 
The kick voltage was set to 2.5 kV. 
The excitation was started from 0.2 sec after the beam injection.
Figure \ref{fig:mountain_excite} shows the mountain plot for the beam excited by the feedback system.

\begin{figure}[h]
\centering
\includegraphics[width=.8\linewidth]{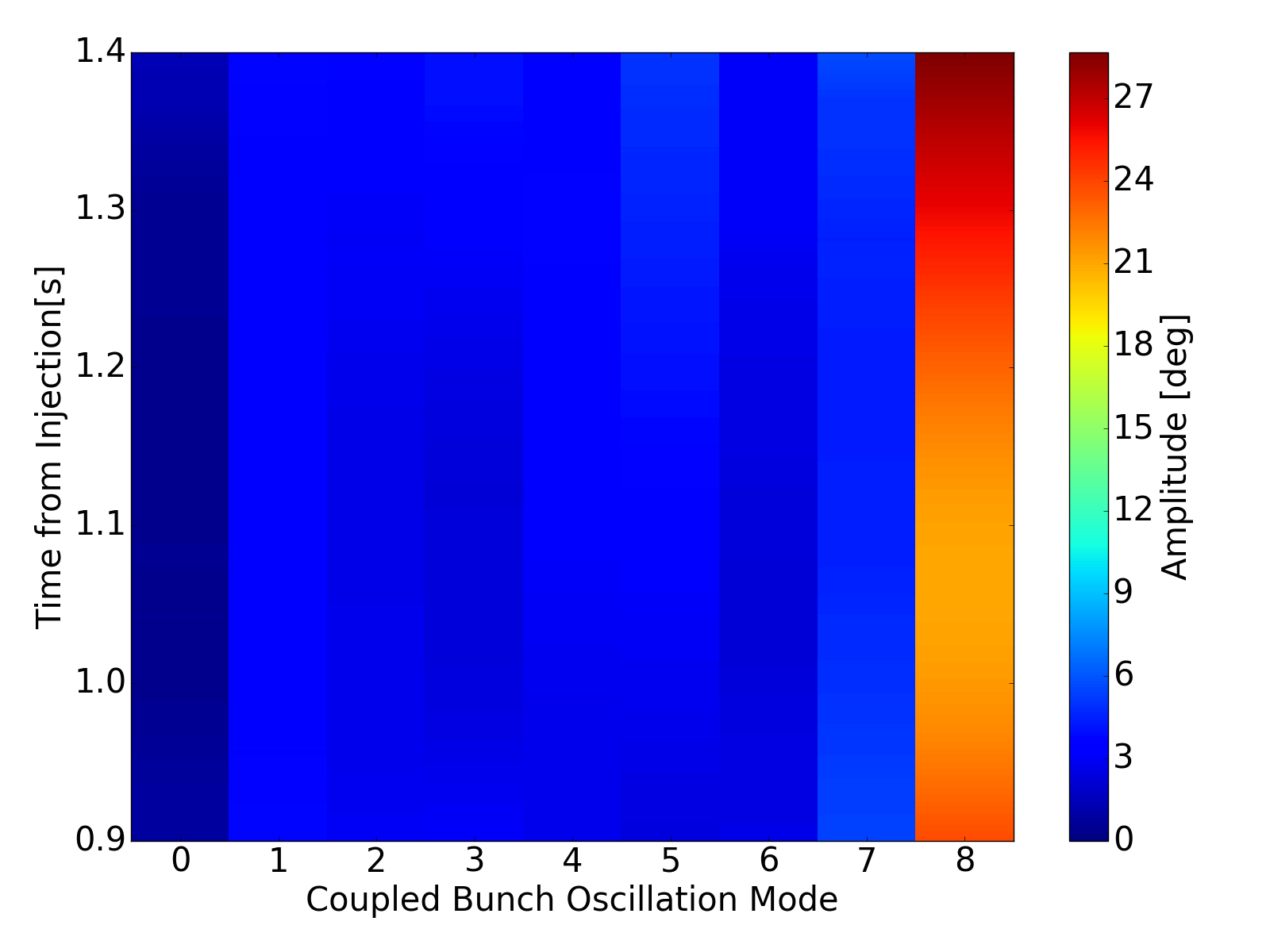}
\caption{The time variation of the oscillation amplitudes of each CB mode based on the bunch centers motion analysis.}
\label{fig:CBOana}
\end{figure}

The oscillation amplitude of each CB mode was obtained by analyzing the motion of bunch centers\cite{Damerau2007,Sugiyama:IBIC2017-TUPCF10}.
The motion of bunch centers was obtained from the analysis of the beam signal recorded by the oscilloscope.
Figure \ref{fig:CBOana} shows the variation of the oscillation amplitudes of each CB mode based on the bunch centers motion analysis.
No significant CB oscillation observed in all CB modes before the excitation,
and only mode $n=8$ has significant growth after the excitation.
This result shows that the CB oscillation of mode $n=8$ was successfully excited.



%
%

%

\subsection{Phase offset LUT adjustment}

The phase frequency response of the feedback system due to delay in the cable and in the SSBF at the feedback processor must be compensated to close the feedback loop.

The phase frequency response of the feedback system against the synchrotron frequency was obtained from the beam excitation measurement.
The phase frequency response of the system was calculated from the difference between the phase of excitation pattern and the phase of the detected oscillation .

\begin{figure}[htb]
\centering
\includegraphics[width=.9\linewidth]{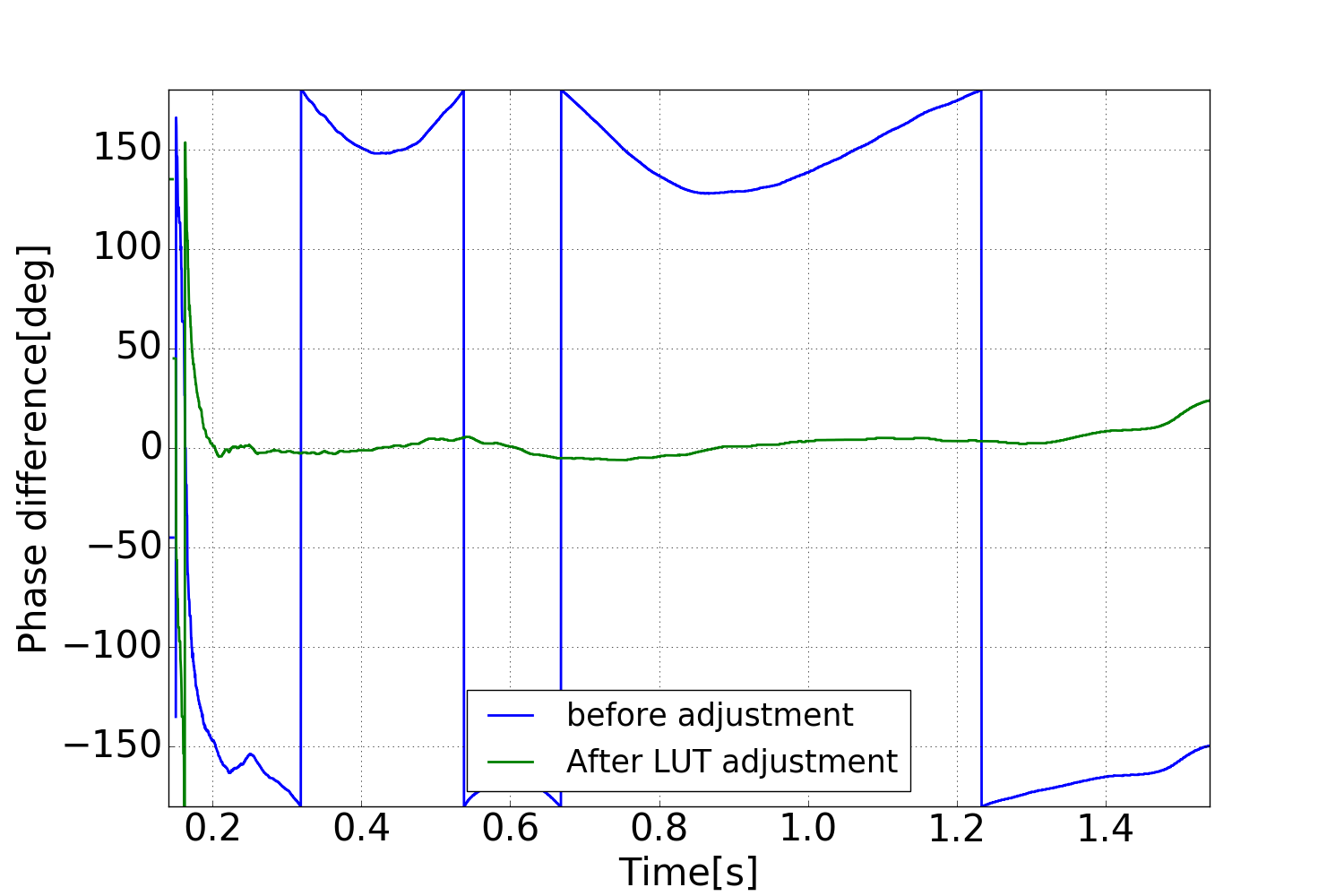}
\caption{Phase adjustment with the phase offset LUT.}
\label{fig:LUT}
\end{figure}

Figure~\ref{fig:LUT} shows the phase difference between the excitation pattern and the detected oscillation phase.
The oscillation of the phase difference suggests that the source of the phase response comes from the characteristics of the logic inside the feedback processor rather than the cable delay. 
After the adjustment of the phase offset LUT, the phase difference was reduced down to less than $\pm$ 5 degree between 0.2 s and 1.4 s from the beam injection.


\subsection{Feedback test}
After the phase LUT adjustment, 
we closed the FB loop and tested the feedback performance of the developed system to damp the CB oscillation.
The feedback performance was tested against the excited beam oscillation.
The same configuration used in the LUT adjustment was used to excite the beam.

\begin{figure}[htbp]
\centering
\includegraphics[width=.9\linewidth]{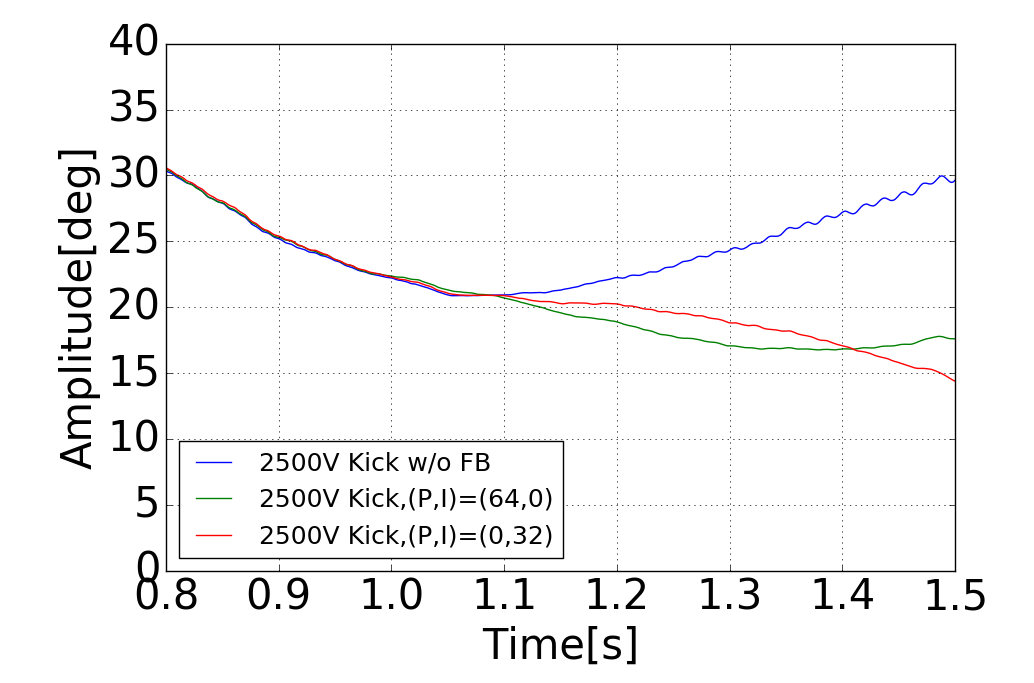}
\caption{Oscillation amplitude of the CB mode $n=8$ for the excited 13kW beam with and without FB.}
\label{fig:PIscan}
\end{figure}

Figure\ref{fig:PIscan} shows the detected CB oscillation amplitude of CB mode $n=8$ for various FB gains. 
Feedback control was enabled after 0.95s from the beam injection.
With P-control or I-control, the CB oscillation was damped down to 50\%.
Better damping is expected with optimized feedback gain since this test was done without feedback gain optimization.

\section{summary and outlook}
We have developed a feedback system for the longitudinal coupled-bunch instabilities in the J-PARC Main Ring.
The feedback performance of the developed system was tested with the beam oscillation excited by the feedback system.
The phase frequency response of the feedback system against the synchrotron frequency was measured with the excited beam oscillation and compensated using a lookup table.
We confirmed the damping of the excited beam oscillation by closing the feedback loop even though the feedback gain was not optimized.
We will investigate the best combination of the feedback gain of P-control and I-control and perform the beam test to suppress the beam oscillation growing without the excitation.

%

\section{ACKNOWLEDGEMENTS}
We would like to thank Heiko Damerau for fruitful discussions on the feedback for the CB instabilities.
We also would like to thank the Mitsubishi Electric TOKKI Systems Corporation for their contribution, including the hardware and the firmware development for the feedback processor. 
Finally, we would like to thank all members of the J-PARC accelerator group for their supports.

\bibliography{IEEEabrv,ForPaper-RT2018}

\begin{thebibliography}{10}
\providecommand{\url}[1]{#1}
\csname url@samestyle\endcsname
\providecommand{\newblock}{\relax}
\providecommand{\bibinfo}[2]{#2}
\providecommand{\BIBentrySTDinterwordspacing}{\spaceskip=0pt\relax}
\providecommand{\BIBentryALTinterwordstretchfactor}{4}
\providecommand{\BIBentryALTinterwordspacing}{\spaceskip=\fontdimen2\font plus
\BIBentryALTinterwordstretchfactor\fontdimen3\font minus
  \fontdimen4\font\relax}
\providecommand{\BIBforeignlanguage}[2]{{%
\expandafter\ifx\csname l@#1\endcsname\relax
\typeout{** WARNING: IEEEtran.bst: No hyphenation pattern has been}%
\typeout{** loaded for the language `#1'. Using the pattern for}%
\typeout{** the default language instead.}%
\else
\language=\csname l@#1\endcsname
\fi
#2}}
\providecommand{\BIBdecl}{\relax}
\BIBdecl

\bibitem{Koseki2012}
\BIBentryALTinterwordspacing
T.~Koseki \emph{et~al.}, ``{Beam commissioning and operation of the J-PARC main
  ring synchrotron},'' \emph{Progress of Theoretical and Experimental Physics},
  vol. 2012, no.~1, p. 2B004, 2012.  Available:
  \url{http://ptep.oxfordjournals.org/content/2012/1/02B004}
\BIBentrySTDinterwordspacing

\bibitem{Nagamiya2012}
\BIBentryALTinterwordspacing
S.~Nagamiya, ``{Introduction to J-PARC},'' \emph{Progress of Theoretical and
  Experimental Physics}, vol. 2012, no.~1, p. 2B001, 2012.  Available:
  \url{http://ptep.oxfordjournals.org/content/2012/1/02B001}
\BIBentrySTDinterwordspacing

\bibitem{Sugiyama2018a}
\BIBentryALTinterwordspacing
Y.~Sugiyama, F.~Tamura, and M.~Yoshii, ``{Longitudinal Mode-by-Mode Feedback
  System for The J-PARC Main Ring},'' in \emph{Conference Record - 21st IEEE
  Real Time Conference, Colonial Williamsburg, USA}, jun 2018.  Available:
  \url{http://arxiv.org/abs/1806.09100}
\BIBentrySTDinterwordspacing

\bibitem{Sugiyama:IBIC2017-TUPCF10}
\BIBentryALTinterwordspacing
Y.~Sugiyama, F.~Tamura, and M.~Yoshii, ``{{M}easurement of the {L}ongitudinal
  {C}oupled {B}unch {I}nstabilities in the {J-PARC} {M}ain {R}ing},'' in
  \emph{Proc. of International Beam Instrumentation Conference 2017 (IBIC'17)},
  2017, pp. 225--228.  Available:
  \url{http://jacow.org/ibic2017/papers/tupcf10.pdf}
\BIBentrySTDinterwordspacing

\bibitem{Pedersen1977}
\BIBentryALTinterwordspacing
F.~Pedersen and F.~Sacherer, ``{Theory and Performance of the Longitudinal
  Active Damping System for the CERN PS Booster},'' \emph{IEEE Transactions on
  Nuclear Science}, vol.~24, no.~3, pp. 1396--1398, jun 1977.  Available:
  \url{http://ieeexplore.ieee.org/document/4328956/}
\BIBentrySTDinterwordspacing

\bibitem{Tamura2013}
\BIBentryALTinterwordspacing
F.~Tamura \emph{et~al.}, ``{Multiharmonic rf feedforward system for
  compensation of beam loading and periodic transient effects in magnetic-alloy
  cavities of a proton synchrotron},'' \emph{Physical Review Special Topics -
  Accelerators and Beams}, vol.~16, no.~5, p. 051002, may 2013.  Available:
  \url{https://link.aps.org/doi/10.1103/PhysRevSTAB.16.051002}
\BIBentrySTDinterwordspacing

\bibitem{Ryoshi2015}
\BIBentryALTinterwordspacing
M.~Ryoshi \emph{et~al.}, ``{MTCA.4 FPGA(Zynq) A/D{\textperiodcentered}D/A
  board},'' in \emph{Proceedings of the 12th Annual Meeting of Particle
  Accelerator Society of Japan}, Tsuruga, 2015, pp. 818--822.  Available:
  \url{https://www.pasj.jp/web_publish/pasj2015/proceedings/PDF/WEP1/WEP116.pdf}
\BIBentrySTDinterwordspacing

\bibitem{Kriegbaum1977}
\BIBentryALTinterwordspacing
B.~Kriegbaum and F.~Pedersen, ``{Electronics for the Longitudinal Active
  Damping System for the CERN PS Booster},'' \emph{IEEE Transactions on Nuclear
  Science}, vol.~24, no.~3, pp. 1695--1697, jun 1977.  Available:
  \url{http://ieeexplore.ieee.org/document/4329055/}
\BIBentrySTDinterwordspacing

\bibitem{Molendijk2017}
\BIBentryALTinterwordspacing
J.~C. Molendijk, ``{Introducing Fixed Frequency Clock Operation on the CERN VXS
  LLRF Platform},'' in \emph{LLRF17}, 2017.  Available:
  \url{https://public.cells.es/workshops/www.llrf2017.org/pdf/Orales/O-22.pdf}
\BIBentrySTDinterwordspacing

\bibitem{Damerau2007}
\BIBentryALTinterwordspacing
H.~Damerau \emph{et~al.}, ``{Longitudinal coupled-bunch instabilities in the
  CERN PS},'' in \emph{2007 IEEE Particle Accelerator Conference (PAC)}.\hskip
  1em plus 0.5em minus 0.4em\relax IEEE, 2007, pp. 4180--4182.  Available:
  \url{http://ieeexplore.ieee.org/document/4439978/}
\BIBentrySTDinterwordspacing

\end{thebibliography}
%
%
%
%
%
%
%
%
%


\end{document}